# INFINITE GROWTH:  A CURSE OR A BLESSING?

## Gennady Shkliarevsky

**Abstract**:  The article discusses the controversy over infinite growth.  It analyzes the two dominant perspectives on economic growth and finds them based on a limited and subjective view of reality.  An examination of the principal aspects of economic growth--economic activity, value, and value creation—helps to understand what fuels economic growth.  The article also discusses the correlations between production, consumption, as well as resources and population dynamics.  The article finds that infinite and exponential economic growth is essential for the survival (conservation) of our economy and civilization.

**Key words**:  Economic growth, technological innovation, Solow, entropy, equilibration, production of disequilibrium, value creation, environmental degradation, and "Malthusian Loop."

## Introduction:  To Grow or Not to Grow

There is a well-known tale related to the game of chess.  The ruler wanted to reward the inventor and asked him to name his price.  The inventor requested what seemed to be a ridiculously small remuneration.  He asked the ruler to place one grain of rice on the first square of the chessboard, then two grains on the second square, four on the third, and then to continue doubling the number of grains for each consecutive square all the way to the last.  The ruler agreed but soon realized that he could not possibly fulfill the request.  By the 30th square the amount of rice that he would have to deliver exceeded all his grain reserves, but this was still way less than the amount that was due for the last square.

The tale is most likely made-up, but it teaches an important lesson:  exponential growth is a powerful force.  In the words attributed to Albert Einstein in another apocryphal story, "compound interest is the most powerful force in the universe."  The subject of both stories is growth, more specifically exponential growth.

The relevance of exponential growth is very broad.  In one form or another, it relates to many spheres of our life, and none of them is more important than the economy.  There are many issues in the world today that are related to the problem of economic growth.  This problem comes up in discussions of climate change, the degradation of the environment, the growing inequality, racial relations, gender discrimination, health, education, and much else.  We hear about the problem of economic growth from politicians, religious leaders, media personalities, and even around dinner tables.  Economic growth seems to be relevant to everything and everyone.

As is common with such important subjects, the problem of economic growth is divisive. It generates heated arguments among scholars, business people, political activists, and common citizens. There are those who unequivocally support economic growth and those who are just as unequivocally opposed to it; and then there is everyone in-between. The point of contention in all these disagreements is whether we can afford an infinite exponential growth or not. For centuries our economic growth has been the source of pride in human genius. Humans believed in economic progress. Infinite growth was the goal they have relentlessly pursued. Today we seem to have reached a critical point in our relentless pursuit of growth. The main question is whether we remain committed to infinite growth or whether we should abandon this commitment.

This article will address the controversy over infinite growth. It will discuss and critically examine the main perspectives on infinite growth that currently dominate the public discourse. It will analyze the theoretical foundation for their respective visions, or what each of these perspectives holds as self-evident truths. These "truths" are not specific to economics and economic systems; they are concepts and notions--such as entropy, chaos, order, and others—that are relevant to many natural sciences, social sciences, and other fields of knowledge. The article will also address specifically economic issues by reviving familiar questions: What is economic activity? What are value and value creation? What is the source of economic growth? Finally, since this article is largely about a disagreement, it will outline a perspective that resolves this controversy and will provide a definitive answer to the fundamental question that is on the mind of all who are involved in it: to grow or not to grow?

## The Controversy Over Infinite Growth

There are many diverse views on growth that are discussed by economists, business leaders, and lay individuals. Despite their diversity, these views come down to two basic perspectives. One perspective rejects infinite growth. It argues that infinite growth will deplete resources, ruin the environment, and may ultimately lead to the destruction of our civilization and the planet. The other perspective sees infinite growth as the only possible way to sustain our economy and civilization. To a non-partisan observer, both perspectives make important points but they also generate confusion and bring a fair share of contradictions into discussions of growth.

### *The Fallacy Argument*

The perspective that argues against infinite growth rests on one categorical statement, according to which nature imposes fundamental constraints on economic growth. Opponents of infinite growth emphasize that there are ultimate limits to the size of human population, production, and consumption that our environment and its resources can support. This statement relies on one fundamental scientific fact about dissipation of energy, or entropy production. As formulated in the Second Law of Thermodynamics, the level of entropy production can never be below zero. According to this law, all resources, including free energy (energy available to do work), will eventually become unusable due to depletion.

Human civilization is a dissipative system. Our economic system consumes low entropy in the environment and produces high entropy. High entropy manifests itself in different ways: either as scarcity of resources, including energy, unavailability of sinks (the capacity of the environment to process waste), or in some other way. But whatever form entropy production takes, high levels of entropy will make our environment very hostile to human life or even completely unsuitable for biological organisms.

The connection between the Second Law of Thermodynamics and economic development emerged at the beginning of the 1970s when Nicholas Georgescu-Roegen published his now famous book *The Entropy Law and the Economic Process*.[1] Since then many new studies on this subject have appeared that both support and reject the validity of the connection between entropy production, economics, and sustainability.

Critics of infinite growth point to the inevitable march of entropy production to make an argument that the pursuit of infinite growth is simply a fallacy that disregards fundamental laws of nature.[2] Christopher Jones, a senior sustainability scientist in the Julie Ann Wrigley Global Institute of Sustainability at Arizona State University, writes that the model of infinite growth has a fundamental flaw and is dangerously delusional.[3] Numerous other researchers and environmental activists have variably called infinite growth a fairytale,[4] a pipe dream,[5] a delusion,[6] and a myth[7]--and these are just some of the milder characterizations. John Cairns is a typical representative of the anti-growth perspective. In his article characteristically entitled "The Age of Transition to Sustainability: The End of the Exponential Growth Period" Cairns writes:

> Human technology, creativity, and ingenuity may modify natural laws, but cannot be used to repeal them. Attempts to maintain the recent exponential growth of the human population, affluence, and artifacts cannot continue forever, or probably even for another century.[8]

Cairns sees only one option: to end exponential growth.[9]

Despite their categorical pronouncements, however, the position of the opponents of infinite growth is not unproblematic. Although they argue that entropy production cannot be stopped and the demise of humanity is inevitable, they also make every effort to convince others that the policies they propose and recommendations they make can work and, if not to avert the inevitable consequences completely, at least delay them and make them more tolerable. Cairns, for example, does not see that ending growth will ultimately save human civilization and its environment, but he is perfectly comfortable with what he sees as an opportunity to "improve the quality of life for future generations of humans and the other life forms, an opportunity that is far greater than possible with present unsustainable practices."[10]

One can draw analogy with the medical profession. Doctors are certainly not trying to convince us there we can live forever, but they do try to prolong our advance to the inevitable end and make it less painful. The analogy works but only to a degree. The policies and recommendations made by the opponents of infinite growth are not without pain of their own. Reducing population and limiting production and consumption will certainly inflict pain, particularly for the average person. What do we have to do to drastically reduce the number of people who live on our planet? What would we have to endure without many modern conveniences to which we have gotten so accustomed? What do we do about those who are currently employed in industries that will have to disappear? These are certainly important questions for which the opponents of growth have no clear answer.

Cairns, for example, recognizes that enforcing limits on population will be painful and most people would oppose them. However, he argues, " . . . if activities that decrease population are not chosen, nature will do so indiscriminately."[11] Melissa Mayer also points to numerous adverse consequences of population growth.[12] The ambiguity of these answers increases tensions and conflicts that do not appear to have any solution other than political enforcement.

*The Technological Panacea*

Supporters of the pro-growth perspective are just as forceful and categorical in their arguments as their opponents. They view the anti-growth stance as profoundly "anti-scientific, anti-empirical, and anti-rational." In a derisive comment, one advocate of growth characterizes this stance as "ancient superstition embodied in an ancient 'argument from incredulity,' prettied up with some modern 'save the planet' lipstick."[13] According to another commentator, there is no alternative to huge increases in economic growth. He writes:

> Contrary to our strong intuitions, infinite economic growth is both
> logically and practically possible on our finite planet, and utterly
> necessary if we want to provide every person on Earth with a decent
> lifestyle. We should not be taken in by prophecies to the contrary, which
> have come and gone since the beginning of human progress itself.[14]

A message that is typical for the proponents of growth is short and curt: "Our baseline expectation should be no less than exponential growth."[15] For Tim Harford, the author of *The Undercover Economist Strikes Back: How to Run--or Ruin--an Economy*, the promise of growth meets the expectation of every generation to live better than its parents' generation: "More growth is better, period."[16] Max Roser, another critic of the anti-growth perspective, the goods and services that we all need

> . . . are not just there--they need to be produced--and growth means that
> their quality and quantity increases. Good health, a place to live, access to education, nutrition, social connections, respect, peace, human rights, a healthy environment, happiness.[17]

The modern pro-growth perspective goes back to the 1950s when the economy in America and the world was rapidly expanding, promising an even better outlook for the future. Robert M. Solow, one of the early expounders of this perspective, has done more than any other thinker to shape growth theory. In several articles that he published in 1956 and 1957 Solow argued that technological change and capital aggregation were two major factors that ensured future growth.[18] One should note that Solow did not specifically refer to infinite growth. However, he also did not dissociate himself from others who made such inferences based on his arguments.[19] The perspective that has eventually been shaped by those who have followed in Solow's steps unabashedly uses the term "infinite growth." A steady flow of contributions that popularize the pro-growth perspective continues to this day and makes this perspective one of the dominant factors in today's economic theory and practice, enjoying particular popularity in the business community and among economists.[20] Many business leaders and economists in China—the greatest economic juggernaut in recent times—see technological innovation to be at the heart of China's economic success story. Analyzing the economic performance of several Chinese provinces that fuel much of the growth in China, Ying Liu and her co-authors conclude that their success has been a result of constant technological innovation. They recommend to "elevate and celebrate" the model of growth through technological innovation in these provinces as the exemplary model for the rest of the Chinese economy.[21]

At the heart of the pro-growth perspective lies the conviction that natural resources and consumption are not the primary factors that fuel economic progress. According to the proponents of the growth model, economic progress is largely due to new ideas, knowledge creation, innovation, and the resulting capital accumulation. Philippe Aghion and Peter Howitt have detailed this view in their piece entitled "Some Thoughts on Capital Accumulation, Innovation, and Growth."[22] For Matthew Johnston, a prolific contributor to *Investopedi*a, "economic growth results from increasing knowledge and applying it to the world to do things better, to improve human welfare"; and the best thing, in his view, is that this growth does not depend on consumption or resources. He and many who share his views see arguments against growth as a profoundly pessimistic admission that "we have already attained all usable knowledge, there is nothing more to learn" and that "increases in knowledge must come to a stop, now or soon"[23]—a view that they find unproven and unacceptable.

The pro-growth perspective does not see any natural limits to the production of knowledge and, consequently, to economic progress that relies on new knowledge and ideas. This perspective largely owes its existence to the view that entropy is irrelevant to knowledge production.[24] Jeffrey Young is in many ways a typical representative of the pro-growth paradigm. He fully subscribes to economic and technological determinism. For him, the problem of growth is primarily an economic (read market) and technological problem. He sees the solution of this problem in technological innovation and market mechanism. In his view, the earth is an open system that imports energy from the cosmos. Consequently, Young has no concerns about energy shortages. His major, if not the only, preoccupation is scarcity of material resources. And this problem, according to Young, can be addressed within the current economic structure aided by technology. The market mechanism—mostly resource pricing—provides the solution. As he argues, "[i]n principal economic models . . . prices which signal relative resource scarcities are sufficient [to resolve the problem of scarcity]."[25] Recovery and recycling have an important but mostly complementary role in his overall scheme.[26]

The proponents of infinite growth do not see entropy as presenting a serious limitation to economic progress. In their view, mental operations are not subject to the laws of entropy and, consequently, knowledge and ideas can significantly reduce and even reverse the production of entropy.[27] Resource scarcity is not relevant to advanced economies that rely on knowledge.[28] Growth advocates point out that "in the advanced economies per capita mineral consumption and energy consumption are both falling" and while their GDP steadily grows, "the portion of GDP represented by material consumption is becoming ever smaller."[29]

*Critique of the Dominant Perspectives on Growth*

The two sides in the controversy over growth use very different principles for organizing their respective theoretical perspectives. Entropy is the central concept for the opponents of infinite growth; they organize their perspective around entropy production. The distinction between mental and material production that has its roots in traditional dualism is the main focus of the proponents of growth. The two organizing principles cannot be more different, so different in fact that the two sides often appear to be talking past each other.

Entropy is indeed a very important part of reality. This fact is indisputable. However, one can take exception from viewing entropy as ontologically primary and entropy production as the process that defines reality. In popular opinion, entropy has a bad reputation. Many see entropy leading ultimately to what is called "thermal death" that is commonly associated with the destruction of the universe. This unfavorable view comes as a result of poor understanding of the important and complex role that entropy production plays in our universe. First of all, entropy production does not destroy anything. It simply dissipates energy from regions where energy density is high to regions where it is low. "Thermal death" does not mean that energy is destroyed. Even if our universe ever reaches this stage, which is a big if, energy will not disappear; "thermal death" simply means that the distribution of energy density throughout our universe will be even and no free energy—energy that can be used to perform work—will be available. In an important way, entropy is not about destruction; it is about conservation.

Entropy production is a form of equilibration. However, the idea that equilibration produces equilibrium reflects a very limited understanding of the effects of equilibration. As a form of equilibration, energy dissipation equalizes energy levels and makes possible for various entities to form permanent bonds, thus creating new combinations that offer new possibilities and degrees of freedom. Such combinations represent new and more powerful levels of organization with more possibilities, more degrees of freedom, and access to new resources. Power differential is about disequilibrium. Thus energy dissipation, or equilibration, produces equilibrium at one level of organization but it also creates disequilibrium by generating a new and more powerful level of organization. As has been argued elsewhere, the two aspects of this process—equilibration and the production of disequilibrium--are closely interrelated and are always in balance.[30]

Many have made a theoretical argument that reality does not show any preference for either equilibration or the production of disequilibrium.[31] Reality, they argue, is constantly in the process of evolution. It is a dynamic system; and as all dynamic systems, it is never either in a state of equilibrium or in a state of disequilibrium, never random or ordered. Dynamic systems are always in a state best characterized as being at "the edge of chaos"—a phrase coined by mathematician Doyne Farmer and popularized by Stuart Kauffman.[32]

There is much empirical evidence that supports this theoretical view. Indeed, one can find many examples when energy dissipation leads to the emergence of new levels and forms of organization of reality. These new levels and forms of organization create new energy flows in a constant cycle of transformation. For example, we can observe a constant cycle of change in our universe from states of reality associated with radiant energy (stars and galaxies) to states of reality associated with non-radiant energy (black holes, gravity, dark energy and matter, and others). Equilibration and the production of disequilibrium are integral to this cycle.

Astrophysicist Manasse Mbonye, for example, observes that "the universe is always in search of a dynamical equilibrium,"[33] which indicates the presence of the interplay between states of equilibrium and disequilibrium. Even though the currently dominant cosmological theory suggests that our universe emerged from some primordial state dominated by disequilibrium—the so-called Big Bang—this theory does not prohibit a possibility that this disequilibrium is not a product of equilibration at some other non-radiant energy level.[34] Therefore, there is little reason to think that at some point our universe was out of balance between equilibration and the production of disequilibrium.

The conclusion is that neither equilibration nor the production of disequilibrium is ontologically primary. One can see entropy as primary only if the process of which entropy production, or equilibration, is only one aspect is not central to one's field of vision.

The main problem of the anti-growth perspective is that it assigns primacy to entropy and, consequently, to equilibration. There is no rational justification for this choice and there is plenty of empirical evidence that contradicts it. Therefore, the organizing principle of the anti-growth is a result of a limited, subjective, and ultimately arbitrary view of reality, which makes the entire anti-growth perspective limited, subjective, and arbitrary. One cannot rely on its conclusions and generalizations to guide our economic policies. The subjective nature of the anti-growth perspective makes it vulnerable to criticism. As a result, this perspective cannot play a constructive role in resolving the controversy over infinite growth. It only deepens this controversy and impedes its resolution.

The distinction between mental and material production as ontologically different and separate is central to advocates of infinite growth. This distinction has its roots in the Enlightenment tradition and is most succinctly expressed by Descartes who saw res cognitans and res extensa as ontologically different from each other. This distinction constitutes the basis for the separation between the subject and the object that is canonical for those who follow in the Enlightenment tradition, the proponents of growth among them. This dualism justifies their exemption of mental production from the ravages of entropy and supports their claims of the possibility of infinite growth. The proponents of infinite growth argue that since mental operations are fundamentally different from physical operations and do not involve dissipation, the law of entropy does not apply to production of knowledge and ideas. If the law of entropy does not apply, there is no limitation to mental production; and since economic growth relies primarily on the production of new knowledge and ideas and this production has no limitations, economic growth also does not have any limits and can be infinite.

The current pro-growth perspective is vulnerable on several points. First, the distinction the proponents of this perspective make mental and physical operations has no rational justification. This distinction goes back to the traditional differentiation between the subject and the object that originates in the Enlightenment tradition and that the pro-growth advocates have accepted uncritically. What we call "subject" and "object" are not ontological entities; they are our mental constructs—mere analytical categories in the way we conceptualize our relationship with reality. As analytical categories they represent the distinction we make between our mental constructs and our capacity to consciously reflect on these constructs and their interrelationship. Both "subject" and "object" have the same source: our mental operations. Changes in the way we view reality are changes that occur in our mind. These changes ultimately reflect new connections among neurons and new configurations of neural circuits that constitute levels of mental organization. In other words, the same process changes the "subject" and the "object." Therefore, assigning ontological significance to this distinction is utterly unwarranted.

Mental operations originate in physical activities. As Jean Piaget has shown in a number of his studies,[35] interactions among functional sensory-motor operations create permanent mental images that open the path to mental operations. In other words, our mental operations originate in physical operations. As any other physical operation, sensory-motor operations generate entropy. Entropy is integral to the process that creates our mental functions. Therefore, entropy production cannot be irrelevant to mental operations; it is totally relevant.

The second point is that the understanding of entropy and entropy production by the proponents of growth is limited. Entropy production is about depletion of resources and resources come in very different forms.  A minor point is that mental operations do involve energy exchanges and, therefore, produce entropy in the most commonly recognized form as energy dissipation.  However, this form of entropy production is not the only and not even the most important one that is involved in mental operations.  Energy exchanges in mental operations are relatively small and consequently the production of entropy is not particularly significant.  Depletion of resources takes another characteristic form specific to mental operations.

Human mind is ultimately a product of the evolution.  The process of creation is central to this evolution.  As a product of the evolution, human mind follows the same pattern that it has inherited in the course of the evolution.  The process of creation involves equilibration; and equilibration generates entropy.  Therefore, equilibration involves depletion of resources.

As any other process, equilibration requires resources that sustain it.  Equilibration feeds on disequilibrium.  Therefore, disequilibrium is its principal resource.  As equilibration proceeds at a given level of organization, the availability of this resource at this level diminishes; in other words this resource depletes and the depletion of a resource is what we call entropy production.

Levels of organization consist of a certain number ($n$) of functional operations.  These operations connect with each other and form stable combinations.  Conservation is at the heart of creating these combinations.  Operations are forms of action; staying active is the main condition of conserving action.  The more active an operation is, the better it is conserved.  This condition is as true of mental operations as it is true for non-mental ones.  New combinations offer new possibilities, new degrees of freedom and, consequently, conserve the entities involved in these combinations.  The $n$-number of functional operations that combine with each other creates the $n^2$-number of combinations, which represents an exponential growth.

As equilibration proceeds and the number of new combinations grows, possibilities for new combinations diminish because a finite number $n$ of functional operations can only produce the finite number $n^2$ of combinations.  This growth is exponential and, therefore, the depletion of the resource is also exponential.  Thus, disequilibrium, or the resource that fuels equilibration at a given level of organization, diminishes, or in other words, like any other resource, this resource is used and depletes.  Although the form this depletion takes is different, in the sense of depletion of resources the human mind is no different than any other system in nature.  It does produce more commonly recognized form of entropy as energy dissipation.  However, and more importantly, it also produces its own specific form of depletion of resources, or entropy production.

The depletion of possibilities for creating new combinations (knowledge and ideas) is not absolute.  It takes place at a given level of organization.  Conservation of new combinations requires regulation.  The level of organization that regulates newly created combinations is more powerful than the level of organization that sustains operations involved in these combinations.  This new global level of organization offers more possibilities, more degrees of freedom, and, consequently, is more powerful than the level of organization that sustains local interactions. The emergence of the new and more powerful level of organization represents the rise of disequilibrium.

The emergence of new levels of organization offers new possibilities. A newly emerging level of organization requires stabilization and conservation. The way to conserve this level of organization is to integrate it with the level from which it has emerged. Naturally, equilibration is the process that makes such integration possible. The integration of the two levels leads to the differentiation of the global regulatory level that leads to further equilibration, creation of new combinations, and the rise of new knowledge and ideas. Entropy production is involved in all these stages. It is never absolute. It is always relative to a given level of organization.

There is also empirical evidence that supports the argument that entropy is relevant to mental operations that generate knowledge and new ideas. Many observers have pointed out that the intensity of intellectual production over the course of human history has been uneven. There have been periods—for example, during the 19[th] and the first half of the 20[th] century—when the growth of intellectual production was very robust; and then there were also periods when intellectual production was slow. For example, many have noted that the production of new knowledge in our time is slow. Our civilization faces many problems today but there are very few new ideas and approaches to solve these problems.[36]

In their insightful article "Are Ideas Getting Harder to Find? Nicholas Bloom and his collaborators challenge "a key assumption of many endogenous growth models that a constant number researchers can generate constant exponential growth."[37] Using a metric of "research productivity" that they have devised, they examine a number of areas where growth heavily depends on research and knowledge. One example they provide relates to computing technology and Moore's Law.[38] Their findings show that the number of researchers needed to sustain Moore's Law has increased 18-fold since 1971 and, consequently, productivity per researcher has fallen in the same proportion. They have also looked in other directions to confirm their findings: agricultural production, medical research, mortality and life expectance, new molecular entities, and aggregates for the economy as a whole. Their robust results show that

> Taking the U.S. aggregate number as representative, research productivity falls in half every 13 years—ideas are getting harder and harder to find. Put differently, just to sustain constant growth in GDP per person, the U.S. must double the amount of research effort searching for new ideas every 13 years to offset the increased difficulty of finding new ideas.[39]

The conclusion that research intellectual productivity is exponentially declining everywhere is indisputable and has been widely accepted.[40]

To sum up, the ontological distinction that the proponents of growth make is not valid. Indeed, entropy production is just as relevant to levels of mental organization, as it is relevant to any non-mental level of organization. It may differ in form but not in its substance and pattern. The proponents of infinite growth are right in their claim that the production of new knowledge and ideas is not limited. However, they are right for a wrong reason. They are wrong in placing the production of knowledge into a special category as unaffected by entropy production. Indeed, production of new knowledge and ideas is infinite. However, it is not infinite because it is immune to entropy production, but it is infinite because it produces new levels of organization that is capable of maintaining the level of entropy production at zero.

The proponents of infinite growth adopt dualism as self-evident truth, for which they provide no rational justification or empirical verification. The uncritical adoption of dualism weakens their position and opens their perspective to criticism by the opponents of growth. Thus, the pro-growth perspective also cannot resolve the controversy over growth. Just like its antipode, it only deepens the controversy and makes it more acute.

The confining of mental operations into a special category and their dissociation from the physical realm makes understanding of the creation of new knowledge and ideas impossible. There is no way to establish control and manage this process without understanding it; and there is no way achieve infinite knowledge creation without conscious control and management.

In conclusion, the above analysis shows that the two dominant perspectives on infinite growth are limited, subjective, and ultimately arbitrary. They have different foundations and, for this reason, largely talk past each other. The anti-growth perspective uses entropy as its organizing principle, while the production of knowledge and new ideas is central for the pro-growth perspective. Neither one, nor the other subjects its organizing principles to a critical analysis. Neither proponents nor opponents of infinite growth offer rational justification or empirical evidence in support of their respective perspectives. There is simply no way that these two perspectives can resolve the controversy over infinite growth.

**Rendering Growth Intelligible**

A solution of the problem of growth is impossible without an understanding of the process of economic growth. Rendering the process of growth intelligible requires the creation of a frame of vision that is broad enough to include all these points and insights as its particular cases, i.e., cases that are true under specific conditions or assumptions. This article will undertake this task in the pages that follow.

*Economic Activity, Value Creation, and Growth: Problems of Definition*

The problem of growth is multi-faceted. Its solution requires a clear understanding of the most important aspects of economic growth; and that is where confusion often starts. Some of the aspects of growth may appear to be fairly straightforward and unproblematic. Yet on close examination they prove to be challenging. Without clarifying these aspects, a solid understanding of economic growth, its source, and its importance for economic systems will be difficult, if not impossible.

As is not uncommon, problems begin with definitions--not lack of definitions (on the contrary, definitions are plentiful) but their quality. This fact is true for economics as it is true in many other disciplines. What often passes for a definition of vital aspects of economic growth, and even a good definition, may be clear but trivial and uninteresting, opening few possibilities for exploring the subject. Reading such definition one gets the impression that everyone understands what it is all about, that no further explanations are needed, and that one should, in fact, feel embarrassed to ask for explanations.

Economic activity is a good starting point in examining the current definitions. Most definitions of economic activity are largely descriptive than explanatory. They simply describe the range of actions relevant to economy. Here is a popular one supplied d by Google:

> Economic activity is the activity of making, providing, purchasing, or selling goods or services. Any action that involves producing, distributing, or consuming products or services is an economic activity. Economic activities exist at all levels within a society.[41]

The problem with this definition is that it does not really explain what is most essential about economic activity. It says little, for example, about what motivates economic activities. Other definitions include such explanations that are confusing and contradictory. For example, one definition states that economic activity "is spurred by production which uses natural resources, labor and capital,"[42] while another explains that "the central purpose of economic activity is the production of goods and services to satisfy needs and wants."[43] These definitions certainly leave one to wonder whether economic activities create needs or needs motivate economic activities.

Lengthier, and fancier, definitions do not necessarily offer more explanatory power. The following definition certainly overloads one's mental circuits but hardly says more than much simpler ones.

> Economic activity is the process by which the stock of resources or stock of capital produces a flow of output of goods and services that people utilize in partial satisfaction of their unlimited wants.[44]

There are, however, some definitions that are clearer than others. Here is one that every businessperson will understand:

> Business begins with value creation. It is the purpose of the institution: to create and deliver value in an efficient enough way that it will generate profit after cost.  Because value creation is the starting point for all businesses, successful or not, it's a fundamental concept to understand.[45]

The definition clearly states that  economic activities are about value creation.  However, its reference to value creation raises questions in the reader's mind as to what constitutes value and how it is created.  The definition offers no clue as to the meaning of this important concept.[46]

If one wants to get confused, looking for a good definition of value and value creation offers one of the best chances.   Definitions of value are numerous, often verbose, diverse, and very different.  The more definitions one reads, the more one falls into a state of confusion, not clarity.

One definition states that economic value is the measurement of the benefit derived from a good or service," but it can also be a "maximum price or amount of money that someone is willing to pay for a good or service."[47]  So, is value same as price?  *Market Business News* explains:  "Economic value therefore refers to the highest amount a consumer is willing to pay for a product or service."[48] In his learned and entertaining volume *The Origin of Wealth:  The Radical Remaking of Economics and What It Means for Business and Society*, Eric Beinhocker tries to provide a very rigorous explanation of value creation.  In his rendition, value creation involves three conditions:  irreversibility (economic transformations should be thermodynamically irreversible); value creation should reduce entropy locally but increase it globally; and fitness (value should fit human purposes).[49]  This definition may be in many ways informative but it still leaves one wondering about the source of value creation.

The number of various theories of value and value creation is staggering.  There is the theory of intrinsic value, the labor theory of value, the monetary theory of value, the power theory of value, the subjective theory of value and marginalism.  There are theorists who explain value creation by social processes such as the division of labor.[50] These are just a few handy attempts to explain value and value creation; there are many others.   Each offers its own version of how value is created.  Some argue that land plays an important role in the creation of value; others maintain that labor creates value, still others assert that value is a result of valuation by a consumer, or that value is produced by social processes, such as the division of labor.  This diversity of definitions and explanations leaves one in confusion as to what or who creates value and how.  The only certainty one gets from this variety is that value is created and that value creation involves new properties.[51]  Therefore, one can conclude that there must be a process that creates new properties, but none of the theorist discusses what such process may look like.

Nitpicking about definitions is not gratuitous.  Knowing what economic activities and value creation are about is essential for understanding economic growth and its source.  Most of the above definitions are not particularly helpful.  There are only two constructive points that emerge from them.  One is that economic activities are about value and that value is created.  Even the theory that claims that value is intrinsic—that is, inherent in the object—still argues that value is a result of the process of valuation. These two points are a good start for discussing growth.

Most current theories of growth are remarkably vague on what constitutes the source of growth. The most popular and widespread view is that growth has its roots in the irrational desire to have more of everything. Elias Canetti, one of the early contributors on the subject, saw the source of growth in a transhistorical "will to grow," which reminds one of the Nietzschean "will to power." In a similar vein, Gareth Dale sees the "desire to accumulate goods, the drive for economic growth, the wish for prosperity" as "innate to human social being."[52] According to Dale, this almost religions desire for growth is instilled in every crevice of our society:

> If there is now one faith, it is faith in production, the modern frenzy of increase; and all the peoples of the world are succumbing to it one after the other . . . Every factory is a unit serving the same cult. What is new is the acceleration of the process. What in former days was generation and increase of expectancy, directed towards rain or corn, . . . has today become production itself. A straight line runs from the rain dance to the Nasdaq.[53]

The French economist Daniel Cohen agrees:

> Economic growth is the religion of the modern world, the elixir that eases the pain of social conflicts, the promise of indefinite progress. It offers a solution to the everyday drama of human life, to wanting what we don't have.[54]

In his well-received volume characteristically entitled *The Infinite Desire for Growth*, Cohen writes in response to his own rhetorical question:

> Why do human beings constantly want to escape their condition? It is an impenetrable question, with which psychoanalysts, anthropologists, and economists have sought to come to terms, each in their own words. But the essential can be summed up in a formula: human desire is profoundly malleable, influenced by the social circumstances in which it finds expression. That makes it insatiable, infinite.[55]

The analysis of the dominant theories, definitions, and explanations related to economic growth shows that they make important points. While these points provide valuable insights, these insights do not amount to an understanding of growth that would be inclusive and yet concise. They ultimately do not render this process intelligible. This task requires bringing the valuable points they make together in a common frame of vision that would be broad enough to include all these points as particular cases, that is, cases that are true under specific conditions or assumptions.

Before constructing such frame, a brief summary of the results of the preceding analysis of current theories, definitions, and explanations is in order.  Here are the main points that emerge:

1. Although many researchers state as a fact that knowledge and ideas advance economic growth, they see the ultimate source of economic growth in the irrational sphere.  As many theorists maintain, an irrational desire to have more of everything ultimately fuels economic growth.  Knowledge and ideas appear to be complementary factors.

2. This irrational desire has no limits.  Consequently, it is capable of generating growth indefinitely.  Therefore, an infinite growth is possible.

3. Growth results from economic activities.  Economic activities are about value, and value is created.

4. The fact that value is created indicates that value does not exist prior to its creation.  In other words, value represents a radical novelty, that is, the emergence of new properties that have not existed prior to their emergence.

5. Economic production depletes resources (both human and material).  In other words, it produces entropy.

6. The rate of depletion is exponential.  As the study by Nicholas Bloom and his collaborators shows, maintaining a linear rate of economic growth requires an exponential increase in the number of researchers, which means that their productivity per person declines exponentially.  The human capacity to produce is a resource and this resource declines exponentially.  This increase indicates an exponential decline in productivity—another indication of depletion.[56]  The conclusion then follows that if productivity increases exponentially, the increase in researchers would not be necessary.  One can apply the same logic to population growth that also follows the exponential power law.  Obviously, an exponential decline in productivity per individual requires an exponential increase in producers.

*Economic Growth and the Process of Creation*

As has been repeatedly stressed, all economic theories agree that economic growth is about value.  They also agree that value is about radically new properties, i.e., properties that have had no existence prior to their emergence.  In other words, these properties have been created and, therefore, there must be a process that creates these new properties.  Yet none of the current economic theories discusses the process that creates radical novelties.

This omission becomes even more curious once one realizes the full importance of this process. It transcends the boundaries of economics and extends to many other parts of reality. We can see its workings at all levels of organization in our universe: from particle, atoms, and molecules, to stars, planets, and galaxies, to life, humans, and civilization. The ubiquity of the process of creation suggests that it is fundamental to the very nature of our universe.[57]

Our universe is unique: it is all there is. There is nothing outside the universe; in fact, the universe has no outside. As there is nothing outside our universe, nothing can come into it and nothing can disappear from it because there is nowhere to disappear. Consequently, everything must be conserved. Conservation originates in the uniqueness of our universe and is essential to its existence.

Conservation requires resources; and resources are always limited. Therefore, access to new resources is the only path to conservation. Since resources are limited, access to new resources can only come with new possibilities and degrees of freedom that can only be a result of new and increasingly more powerful levels of organization. When particles combine into atoms or when atoms form molecules, they acquire a broader range of possibilities, or degrees of freedom, and give rise to new properties. When hydrogen and oxygen combine they create a new molecule—water. The properties of water are different from those of oxygen and hydrogen. The temperature of the liquid state of water, for example, is much higher than temperature of the liquid states of hydrogen or oxygen.

New possibilities offer access to new resources that are essential for conservation. Simple molecules can combine into complex molecules (for example, organic molecules), which expands their possibilities. Organic molecules allow even more diverse combinations that create new and even more powerful levels of organization that are capable of sustaining life. The emergence of new and increasingly more powerful levels of organization is the very essence of the evolution. For what is the evolution if not a succession of cascading increasingly more powerful levels of organization nested in each other *matryoshka* style?[58]

Yet despite this singular importance of the process of creation, we know very little about it and study it even less.[59] Moreover, we generally do not regard this process as intelligible and accessible to human understanding. Margaret Boden, one of the pre-eminent researchers in the field of creativity, draws the following conclusion in her influential book on creativity:

> Our ignorance of our own creativity is very great. We are not aware of all the structural constraints involved in particular domains, still less of the ways in which they can be creatively transformed. We use creative heuristics, but know very little about what they are or how they work. If we do have any sense of these matters, it is very likely tacit rather than explicit: many people can be surprised by a novel harmony, but relatively few can explicitly predict even a plagal cadence.[60]

All current economic theories agree that new knowledge and ideas generated by the human mind are the primary source of economic growth.[61]  The human mind has emerged in the course of the evolution that is driven by the process of creation.  As a result, the human mind has inherited the properties of the process that led to its emergence.  Therefore, the way that the human mind operates has the same pattern as the universal process of creation.  One sees in its operation the same sequence of conservation, creation of new combinations, and the emergence of new and increasingly more powerful levels of organization that give rise to new properties (new knowledge and ideas).  It obeys the same pattern of exponential growth due to the same operational combinatorics that characterizes the universal process of creation.

Indeed, mental constructs are our subjective creations.  However, the process that we use to create them is not.  Humans have not created this process but rather inherited it in the course of the evolution that preceded the rise of humanity.  This process transcends human existence.  It is objective and does not depend on mental constructs created by humans.  Humans merely use this process.

As has been pointed out earlier, there is the fundamental connection between the process of creation and conservation.  Therefore, economic growth is also primarily and objectively about conservation of the existing levels of organization.  Certainly, economic growth has ancillary effects:  expanding availability of goods and services, increased consumption, improved quality of life, and others.  Indeed, these effects may serve as motivations in pursuing economic growth.  However, what ultimately drives economic growth is conservation of the existing levels of organization and, first and foremost, our current levels of mental organization—the source of our knowledge and ideas.

New and more powerful levels of organization give rise to new knowledge and ideas.  How does the process of creation generate new levels of organization?  Each level of organization sustains a number of operations.  Operations are a form of action; and conservation of action requires its enactment.  Conservation of functional operations requires their activation:  the more they are activated, the more stable they are and the better they are conserved.  Activation is a resource for functional operations.  New and stable connections with each other activate operations more often and thus conserve them better.

Just like the production of new properties in equilibration of physical entities, the equilibration of mental operations gives rise to new levels of organization that offer new possibilities, new degrees of freedom, and new properties.  The level of organization that sustains mental operations and consciousness originates in physical activities.  As Jean Piaget has shown, combinations of sensory-motor functions (visual, audio, tactile, olfactory, and gustatory) generate mental images that open the path to mental operations.[62]  Sensory-motor functions are operations that connect organisms with external reality.  These functions make possible using external reality and its resources to sustain the organism.  The operation of sensory-motor functions is in this respect essentially a form of economic production that also derives resources from the environment to sustain human life.

Conservation of human systems, just like conservation of other systems, requires integration of all levels of organization. Mental operations represent the global level of organization that regulates physical functions from which this level has emerged. The conservation of this level of organization requires its equilibration with the level of physical operations, which means that novelties created at the level of mental operations must be translated and assimilated at the level that sustains physical functions. The connection between mental operations and physical operations goes both ways. Both are part of human systems where they represent different levels of organization. The translation of radically new knowledge and ideas into economic production conserves the entire system. Indeed, specific benefits that improve human life will also accrue from such integration. But they are not the only or even the most important result of such integration; the most important result is the conservation of the entire human system.

*Economic Growth and Its Correlatives*

The opponents of economic growth insist that the survival of our civilization vitally depends on limiting production, constraining consumption, and reducing population. The correlations between growth, production, consumption, and population are central to this argument. Anti-growth theorists are not the first to establish these correlations. Back in the 18[th] century the famous British economist Thomas Robert Malthus formulated these correlations in a book entitled *An Essay on the Principle of Population.*[63]

Malthus argues that linear production growth causes an exponential increase in population. The discrepancy between the two rates of growth leads to overpopulation. When resources necessary to support the growing number of people exceed economic production capacity, a reduction of the population becomes inescapable. In Malthus's view, wars and pestilence are inevitable results of overpopulation. They reduce the number of people and restore the balance between the existing level of production and the number of people this production supports. Malthus's formulation has become known as the "Malthusian Loop" or the "Malthusian Trap."

There has been much criticism of Malthus and his theory and there is no need to rehearse this criticism here again.[64] Two points, however, are in order for the purposes of this article. Malthus has never adequately explained why an exponential population growth is an inevitable response to production increase. After all, instead of increasing birth rate, people can very well decide to concentrate on benefiting from growth and improving their conditions. There is nothing inevitable about this response from the population. Also, Malthus has never adequately explained what causes population to grow exponentially when the means of sustaining the growing population does not increase at the same rate. Wouldn't it be wise for people to adjust the two rates and avoid future calamities?

The correlations between production, consumption and population are also central for the contemporary anti-growth theorists. They have largely adopted it, albeit with significant modifications. They see growth of production and consumption as the ultimate threat to the terrestrial environment and resources. Population growth also remains part of the overall equation as a compounding factor. Although, unlike Malthus, the modern anti-growth theorists view population growth as an independent variable,[65] rather than a product of economic advances, they still see the need to reduce population since excess of people, in their view, contributes to depletion of resources and degradation of the environment.[66]

The arguments of the opponents of growth raise objections. For example, as has been shown earlier, economic growth does not necessarily has to deplete resources; on the contrary, it can reduce consumption of resources per individual.[67] Also, critics of growth also do not explain the presence of the exponential index in their calculations of the depletion of resources. Likewise, they offer no explanation why population grows inexorably and why this growth is also exponential.

The proponents of growth also recognize the correlations between production, consumption, depletion of resources, and population growth. However, they believe that production can be effectively divorced "as much as the laws of nature allow" from inputs of natural resources, thus permitting growth without depletion. They emphasize that technological change--not just its rate but also the type—can make this de-coupling possible, thus ensuring growth without environmental degradation.[68] One should also add that that, like anti-growth theorists, they also do not explain the presence of the exponential index in their calculations.

Correlations of production, consumption, depletion of resources, and overpopulation are indicative. They are too consistent to be dismissed as mere coincidences. They indicate a persistent dependency that requires an explanation. The fact that these correlations follow the exponential power law suggests that an exponential increase in economic growth will compensate for the depletion of resources and will offset the need for population increases.

The perspective that centers on the process of creation answers the questions that dominant theories do not answer. It agrees with the dominant economic perspectives in recognizing that new knowledge and ideas are the most important factors affecting economic growth. According to this perspective, new knowledge and ideas arise from new and more powerful levels of organization generated by our mind.

This article emphasizes the role of conservation in the process of creating new levels of organization. Conservation of the operations sustained by a particular level of organization results from their combinations with each other. The equilibration of an $n$-number of operations sustained by this level will generate a new and more powerful level of organization that will sustain the $n^2$-number of operations, thus displaying an exponential growth.

The human mind and consciousness can generate an infinite number of new and increasingly more powerful levels of mental organization; and the increase in power for each level will be exponential; it will follow the power law of $n^2$. There is an obvious connection between the two levels of organization and this relationship goes both ways. Exponential increases in power at the level that sustains mental operations will translate into increases in power at the level of physical operations, or material production, that it regulates.[69] A failure to translate exponential increases at the level of mental organization into economic growth threatens the conservation of mental operations and may lead to the disintegration of the economy and the entire human system.

If we fail to maintain exponential growth, we will not be able to gain access to new resources. Consequently, in order to sustain the linear rate of economic production, we have to increase exponentially our use of available resources—both material and human. The exponential increase in the use of resources will result in resource depreciation and environmental degradation.

Contrary to what the opponents of growth argue, cutting production to save resources will offer no relief. The source of the problem is in lack of growth and consequently underutilization of available resources, particularly human resources. The overuse of natural resources comes as a consequence of underutilizing human resources, specifically human capacity to create new and increasingly more powerful level of organization. Real growth is in creating new levels of organization, not simply increasing outputs at a given level of organization. Reducing production at the current level does not provide access to new resources. It merely leads to the non-conservation of the current global level of organization in the human system, which will make the use of resources even more inefficient. The result will be a progressive and exponential depletion of resource and environmental degradation. Only an exponential economic growth offers the solution.

The same pattern is present in the relationship between economic growth and population. People are "the ultimate resource" that we have.[70] A failure to achieve an exponential growth will result in an exponential decline in productivity, or inefficiency in using the human resource. Studies cited earlier show that if we fail to achieve an exponential growth in productivity, we will need to use more people.[71] That is what creates the momentum for population growth; and this growth follows the exponential power law, as Malthus has shown. Considering Julian Simon's point that humanity itself is "the ultimate resource," the conclusion that a stagnant population is the ultimate resource crisis is quite appropriate.[72]

This is to argue that the cause of population growth is inefficient production. Attempts to limit this growth will never succeed because they do not address the main cause that generates exponential population growth—exponential increase in production inefficiency. All efforts to limit the global level of population have failed. There have been numerous summits and conferences over the last several decades that have provided platforms for discussions of population problems.[73] The only result that they produced, as critics charge, were numerous statements to the effect that "reducing population growth was a necessary part of ecologically sustainable development."[74] This singular lack of success has caused much criticism of the entire population science. George Weigel, president of the Ethics and Public Policy Center in Washington, D.C.—one of the harshest critics—has charged that population science has little merit and only produces myths, miscalculations, and bogus predictions.[75]

Population growth is not an independent variable.[76] Its source is production inefficiency that grows exponentially and requires exponential compensations. The continued acceptance of child labor in the developing world as a partial solution to poverty speaks to the shortage of labor due to low productivity.[77] Efforts to use compulsion to reduce population will be extremely unpopular and utterly ineffective. They will simply put more pressure on the economy that will be unable to maintain its current level of economic production. The conclusion is certainly counterintuitive, but the logic of population growth suggests that an efficient use of human resources requires exponential growth in productivity. Such growth is the only way to avoid overpopulation.

Finally, there is one more correlative to economic growth that is high on the anti-growth agenda. It is consumption. The opponents of growth claim that out-of-control consumption drives production to unsustainable levels. Therefore, they advocate reducing consumption as a way to constrain production. Their position is not devoid of ambiguities. They recognize, for example, that there is an urgent need to improve life conditions for underprivileged groups and populations in poor countries. As a way of addressing this problem of the growing economic disparity, they propose a more equitable wealth distribution. In other words, they want to use redistribution of wealth to increase consumption among the underprivileged. Consequently

A close analysis of the position of the opponents of growth reveals ambiguity in their understanding of the relationship between production and consumption. They propose policies designed to reduce consumption as a way of constraining growth. However, they also believe that they can increase consumption among underprivileged groups and at the same time reduce production. In other words, their proposed policies reveal that they hold two different views of the relationship between production and consumption. On one hand, they see them as deeply interrelated: an increase of consumption leads to an increase in production and vice versa. On the other hand, they also believe that they can increase consumption without increasing production. They do not explain why they hold such diametrically opposed views. They apparently do not even see any paradox in the way they view the relationship production from consumption.

Consumption is a form of assimilation, i.e., the inclusion of one entity into functional operations of another. When operations combine, they assimilate each other, or include each other into their functional activities. In a way, they consume each other, or use each other to conserve and sustain their own functions. But, at the same time, this consumption produces new combinations--new entities.

Thus, production and consumption are intimately interrelated. They are both aspects of the process that creates new properties, or what economists call value creation. They are analytical, rather than ontological categories. As the analysis of the process of creation shows, the conservation of a given level of organization creates a new and more powerful level of organization. The emerging level of organization supervenes on the level from which it emerges; in other words, the level of organization that gives rise the new level of organization is a resource that is consumed in the process of creation. One can also represent this relationship between consumption and production as a balance between equilibration and the production of disequilibrium. In this conception, disequilibrium is a resource for equilibration that produces new disequilibrium.

There is another example to illustrate the close relationship between production and consumption. Our sensory-motor functions produce and consume reality at the same time. Our relationship with reality sustained by sensory-motor functions involves both consumption and production. When we observe reality, we organize/produce it according to the structure of our sensory-motor functions; at the same time, when our sensory-motor functions include this organized reality into their operations they consume this organized reality, which conserves sensory-motor functions and ensures their continued operations. Viewing production and consumption through the prism of the process of creation shows their fundamental complementarity. Only when the process of creation is not central to one's frame of vision, production and consumption appear as ontologically separate and independent from each other.

The ambiguity in the approach by the opponents of growth toward the problem of consumption shows that the process of creation is not central to their frame of vision. This fact reflects a more general condition in our economic production that is not organized around the process of creation. A dissociation of consumption and production is not uncommon in our economy that has a strong tendency to underutilize and waste our most important resource—human capacity to create. There are also numerous examples when consumption is wasteful and does not contribute to growth in productivity.

By producing new and increasingly more powerful levels of organization we generate new knowledge and ideas that lead to economic growth. Knowledge, for example, is one important product that does not depreciate. It only appreciates when consumed, as its consumption leads to new and increasingly more powerful levels of organization that give rise to new knowledge and ideas. As Thomas Davenport and Lawrence Prusak have noted, "ideas breed new ideas, and shared knowledge stays with the giver while it enriches the receiver."[78]

Organizing our economic activities around the process of creation will end this tendency to dissociate production from consumption; it will make will make the interrelationship between production and consumption effective and efficient. By complementing each other, they both will be able to grow exponentially and will make possible an exponential growth of our entire economy. One can see the contours of this new economic organization in the comment of Alan Webber who concludes: "In the end, the location of the new economy is not in the technology, be it the microchip or the global telecommunications network. It is in the human mind."[79]

*The New Economy of Exponential Growth*

If economic growth is so important, if the process of creation that fuels economic growth is so powerful as to sustain our entire universe, why do we fail to use it and achieve an infinite exponential economic growth? The reason is due primarily to the fact that we have not embraced the process of creation. We do not understand how it functions and, consequently, cannot control and use it fully to our advantage.

Consciousness is the most distinct feature of human nature. It represents the most powerful level of mental organization. Therefore, if we fail to achieve exponential growth, the reason for this failure can only reside in consciousness.

Just like any other level of organization, the level of organization that sustains consciousness and its operations must be conserved; therefore, consciousness must engage in creating new and increasingly more powerful levels of its organization. Since consciousness is involved, engaging the process of creation cannot be a spontaneous development. It must be a result of conscious efforts and, therefore, it must be intentional. Our consciousness should embrace the process of creation. We should understand how the process of creation operates and we should consciously and intentionally use this process as the main organizing principles of our economic practice.

Many researchers indicate the need to change our economic practice. Maria Sandberg and her co-authors express the conviction of many scientists that "changes in the organization of human society and economy are needed to stop the degradation of the natural environment."[80] One can also see this expressed need in the call for changing managerial and business practices.[81]

The fact that we still fail to embrace and understand the process of creation is no accident. The source of this failure has deep evolutionary roots that I have explained elsewhere.[82] Although this failure has evolutionary roots, it is not irremediable. Our consciousness represents the most powerful level of organization of reality. It is capable of solving any problem even if such problem is due to its own failing.

Contrary to what the opponents of growth maintain, there are no inherent physical obstacles, such as entropy production, to achieving infinite and exponential economic growth. Indeed, according to the Second Law of Thermodynamics, the level of entropy production cannot be less than zero. However, there is nothing in this law that prohibits entropy production to stay at a zero level.

There are some specific reasons why we have so far failed to embrace fully the process of creation. For one thing, we still do not believe that this process is accessible to human understanding. The process of creation operates on inclusion.  Yet the social practice that dominates our civilization has been and remains exclusionary.  We fear differences and seek to shun and suppress them, which smothers the process of creation.  Finally, there is a great deal of institutional inertia and reliance on old and tired dogmas.  All these obstacles should be removed.  By embracing the process of creation and using it as the main organizing principle of our practice, we can achieve exponential growth.

The consequences of not achieving exponential growth are with us.  We live with them.   We deplete our resources, including human resources.  The decline of productivity is exponential and so is the depletion.  The exponential decline in productivity also makes necessary the exponential increase in population.  Reducing population is not a substitute for an exponential growth of productivity and neither does the redistribution of wealth.  Both are irrelevant to economic growth.  They can only make the effects of a lack of growth more serious.[83]

Without infinite exponential economic growth we cannot sustain our civilization.  Due to lack of exponential growth, we cannot maintain the production of entropy at a zero-level.  The inevitable result is a progressive degradation of our environment and an unsustainable growth of the population. As many have already recognized, the continuation of these trends will lead to the destruction of our civilization and our planet.

**Conclusion**

The importance of the problem of growth is hard to overestimate.  The health of our economy and society and the sustainability of our civilization vitally depend on the solution of this problem.  The fact that this problem still remains unsolved is arguably the biggest threat to the survival of humanity.

This article has shown that the survival of our civilization requires the constant creation of new and increasingly more powerful levels of organization.  Each new level of organization new level of organization represents an exponential increase in power.  The article has also explained the reason why this growth should be infinite and exponential.  Here is a recapitulation of the main points:

1.  The conservation of a system, including human systems (such as, for example, civilization), requires regulation.

2.  The level of organization that regulates a particular system is always the most powerful level of organization in this system.  It emerges as a result of the equilibration of all functional operations of this system.  Since the emergence of a new level involves the creation of combinations of operations, power of this level of organization is always exponentially greater than the power of the sum total of all operations in regulates.

3. The regulatory level of organization also requires conservation that can only be provided through the creation of a new level of organization that will also represent an exponential increase in power in comparison to the level it conserves.

4. Thus, the sustaining of our civilization, including its economic system, requires a constant exponential increase in power, or exponential growth.

5. Moreover, since we can and, indeed, must always create a new level of organization that would regulate and conserve our current level, this growth is infinite.

6. A failure to achieve an exponential growth inevitably leads to an exponential decline in productivity, an exponential rate in the depletion of resources, and an exponential population growth.

As this article has shown, the two dominant perspectives on growth do not provide a definitive solution to the problem of growth. On the contrary, they have only deepened the controversy, added to confusion, and made the resolution less likely, if not indeed impossible. They have become part of the problem, rather than part of its solution. Tony Lawson has thus characterized the current state of economics as a discipline:

> I want to argue, however, that the remedies suggested, if in the context sometimes remarkable, are not going to make things sufficiently better. And the reason for this is simply that the problems identified do not represent the real nature of the relevant disorder. Economics is a discipline in some disarray. Most notably its most prestigious branch, "economic theory," appears quite unequal to the tasks of explaining real-world events or informing policy analysis.[84]

There is no need to accept this sad state as inevitable. We need a new perspective on the familiar and broadly discussed economic issue. Economics does not have to be a guessing game. One can agree with Tony Lawson that it "can easily be transformed into a subject that is potentially illuminating, consequential for policy design, and even emancipatory."[85]

This article recognizes that both current perspectives on growth make valid points. One can certainly agree with the recognition of the importance of entropy production emphasized by the opponents of infinite growth. However, one should also acknowledge, as this article does, that their understanding of the role of entropy production is very limited. In their view, entropy production only depletes resources and creates disorder. They do not see that entropy production also produces order. Equilibration that is central to entropy production also generates new and increasingly more powerful levels of organization and, thus, disequilibrium, or the foundation of a new order.

The limited understanding of the role of entropy leads to conclusions and policy recommendations that are wrong-headed and even harmful. They cannot solve the problems we face; they can only lead to a disintegration of our production system and, ultimately, the civilization. The course that the opponents of growth chart will work against the creation of new and increasingly more powerful levels of organization that is essential for sustaining our economy and civilization. This course can only lead to a progressive dismantling of the levels of organization that we have already created. It leads into a dead end.

One can also appreciate the emphasis on human creative capacity made by the proponents of infinite growth. This capacity has certainly served humanity well in the course of its history. However, the proponents of growth base their perspective on the recognition that mental and physical operations are fundamentally different. They do not provide a rational justification or empirical verification for this assertion that ultimately is a reflection of the traditional mind-matter dualism. This emphasis puts the creation of knowledge and ideas into a separate category that, unlike physical activities, is inaccessible to human reason. As a result, the perspective formulated by the pro-growth theorists is not conducive to understanding the process that generates knowledge and ideas—an understanding that is essential for controlling and managing the process of creation and thus enhancing economic growth.

There are also other problems with the current pro-growth perspective. It tends to overemphasize the importance of technological innovation and underestimate the need for a more comprehensive approach that would involve all levels of the human system. The proponents of growth are overly enthusiastic about quick fixes. As the ultimate solution of the current economic problems, digital and communication technology, for example, generates a great deal of unjustified optimism.[86]

Technocratic quick fixes may catch headlines, but, as many researchers recognize, they will not solve the problem of growth. While technological innovations are important, they are not sufficient. Summarizing the current tendency to lionize technocratic utopian solutions, Christian Cancino and his co-authors emphasize a broader intellectual context that new solutions require. They write:

> There is an excessive reliance on technological innovation. They do not understand that technological innovation does not occur in a vacuum. Just as other spheres of intellectual pursuit, it is dependent on the creation of new levels of organization. It is, for example, co-dependent on science . . . Their partial perspectives and lenses generate new knowledge in their own domains, but they are not necessarily compatible and complementary in a more holistic perspective.[87]

Marinko Skare and Małgorzata Poradarochon also point to the insufficiency of technological progress alone. In their view, technological innovation "is only a necessary prerequisite for long-term sustainability but is not sufficient."[88] Mirjam Beltrami, among others, stresses the impact of technology on other spheres of economic practice and sees the need for comprehensive changes.[89]  These are just a few of many who call for a comprehensive approach that will reorganize our economic, and not only economic, practices.

While the article accepts the point about the importance of new technological ideas, it also contends that human capacity to create knowledge and generate new ideas does not stand apart from the natural order but lies fully within it.  By taking this approach, the article stresses the need of using the study of the process of creation in nature in a way that will complement and enrich all current perspectives on economic growth.  The perspective outlined in this article recognizes and accepts the valid points made by both the proponents and the opponents of growth.  It incorporates these points into a broader frame as its particular cases—that is, cases that rely on special assumptions or conditions. The key to such integration is the recognition of the centrality of the process of creation. This article sees the failure to understand this process as a serious flaw in both dominant perspectives.

The central focus and the organizing principle of the inclusive perspective outlined in this article is the process of creation.  The article argues that the failure to grasp the centrality of this process leads to fragmentary and contradictory views of reality that characterizes both dominant perspectives.  By offering a view that is inclusive and comprehensive, the perspective outlined in this article resolves the controversy over infinite growth.  It makes clear that exponential growth is essential for sustaining our civilization and this growth can and must be infinite.  Only by constantly creating new and increasingly more powerful level of organization, we can ensure the survival of human civilization.  This perspective also suggests that we will be well advised to use the process of creation as the main organizing principle of our social practice, including economic production and management.

This article shows that infinite exponential growth will lead to increases in productivity and will reduce the depletion of resources.  It will constrain population growth since there will be no expressed need to use more people to compensate for the inadequate growth. The fact that the exponential index $n^2$ appears in the process of creation, in depletion of resources, in the growth of the population and other phenomena correlated with economic growth is not accidental.  These correlatives are closely related to the process of creation that is fundamental to nature.  Nothing short of an infinite and exponential growth can sustain our civilization.  Any proposal that promises to save our civilization by ending or constraining growth and by reducing production and consumption has no basis in reality; and neither do the proposals that seek to reduce population as a way of saving humanity. Such proposals can only lead to the disintegration of human systems, including our economy.

Infinite and exponential growth is the only path forward toward sustaining human life and civilization.  Such growth is the fulfillment of the destiny bequeathed to the human race by the entire evolution of nature:  to continue creating new and increasingly more powerful levels organization indefinitely, thus shaping and reshaping the future of our civilization and that of the universe.

**ENDNOTES**